# Correctness of an Incremental and Worst-Case Optimal Decision Procedure for Modal Logic with Eventualities

Mark Kaminski and Gert Smolka
Saarland University

February 11, 2011

We present a simple theory explaining the construction and the correctness of an incremental and worst-case optimal decision procedure for modal logic with eventualities. The procedure gives an abstract account of important aspects of Goré and Widmann's PDL prover. Starting from an input formula, the procedure grows a Pratt-style graph tableau until the tableau proves or disproves the satisfiability of the formula. The procedure provides a basis for practical provers since satisfiability and unsatisfiability of formulas can often be determined with small tableaux.

## 1 Introduction

We are interested in practical decision procedures for modal logics with eventualities. The standard logic in this area is PDL [5], which is the propositional fragment of dynamic logic [15, 10], a formal system for reasoning about programs. Fischer and Ladner [5] establish the decidability of PDL with a small model theorem, which states that every formula has a syntactic model obtained over a finite universe now known as Fischer-Ladner closure. They also show that the satisfiability problem of the fragment of PDL just consisting of basic modal logic and eventualities is EXPTIME-hard. Based on Fischer and Ladner's model construction, Pratt [16] gives an elegant decision procedure for PDL that runs in deterministic exponential time. Variants of Pratt's procedure appear with correctness proofs in [10, 3].

Although worst-case optimal, Pratt's [16] procedure is not practical since it starts with all subsets of the Fischer-Ladner closure. Pratt [17] thus devises a more elaborate procedure that employs graph-like structures we call graph tableaux. Given a formula, this procedure constructs a complete graph tableau for the formula and checks whether the tableau contains a model of the formula.

While Pratt's tableau procedure [17] is an improvement over his abstract procedure [16], it is still unsatisfactory since the construction of a complete graph tableau is often too expensive. Goré and Widmann [8, 9, 18] thus devise an incremental method that interleaves the tableau construction with model checking. Their method is realized in an impressive prover [7, 11], which often decides the satisfiability of a formula without building a complete tableau.

In their papers [8, 9], Goré and Widmann report in detail about the algorithms used in their prover but do not include correctness proofs. Correctness proofs appear in Widmann's thesis [18], but they are monolithic and refer to complex algorithms.

In this paper we present a simple theory explaining the construction and the correctness of an incremental and worst-case optimal decision procedure for modal logic with eventualities. The key idea is to have two complementary algorithms called eager and cautious pruning, which check whether a graph tableau certifies the satisfiability or the unsatisfiability of a formula. The decision procedure then incrementally constructs a tableau certificate for the input formula, where eager and cautious pruning guide the expansion of the tableau.

Our theory explains important aspects of Goré and Widmann's [8, 18] method for PDL that are not addressed in their work. For simplicity, we do not treat full PDL but consider basic modal logic with simple eventualities. We expect that our theory extends smoothly to full PDL. Goré and Widmann [9, 18] have extended their method to PDL with converse. The extension to converse is not straightforward and will require a significant update of our theory.

The paper is organized as follows. We first introduce the underlying logic and define a class of syntactic models called demos. We show that a formula is satisfiable if and only if it is supported by a demo. We then introduce abstract pruning, which captures Pratt's abstract decision method and serves as the blueprint for the pruning operations on tableaux. Then we define graph tableaux and establish the subclass of evident tableaux, which certify the satisfiability of formulas. We then consider eager and cautious pruning, two complementary algorithms identifying satisfiable and, respectively, unsatisfiable clauses in tableaux. Now everything is in place for the incremental decision procedure. We conclude with remarks on branching tableau search.

## 2 Formulas and Models

We consider basic modal logic extended with the star modalities $\Diamond^*$ and $\Box^*$. We call this logic K*. **Formulas** are defined with the grammar

$$s ::= p \mid \neg p \mid s \vee s \mid s \wedge s \mid \Diamond s \mid \Box s \mid \Diamond^* s \mid \Box^* s$$

where $p$ ranges over a nonempty set of names called **predicates**. We consider only formulas in negation normal form. Formulas with general negation can be translated in linear time into negation normal form. We write $\Diamond^+ s$ for $\Diamond\Diamond^* s$ and $\Box^+ s$ for $\Box\Box^* s$. An **eventuality** is a formula of the form $\Diamond^* s$ or $\Diamond^+ s$. A **model** $\mathcal{M}$ consists of the following components:

- A nonempty set $|\mathcal{M}|$ of **states**.
- A **transition relation** $\rightarrow_{\mathcal{M}} \subseteq |\mathcal{M}| \times |\mathcal{M}|$.
- A set $\mathcal{M}p \subseteq |\mathcal{M}|$ for every predicate $p$.

We write $\rightarrow_{\mathcal{M}}^*$ for the reflexive and transitive closure of the transition relation. The **satisfaction relation** $\mathcal{M}, w \models s$ between models $\mathcal{M}$, states $w \in |\mathcal{M}|$, and formulas $s$ is defined by induction on $s$:

$$\begin{aligned}
\mathcal{M}, w \models p &\iff w \in \mathcal{M}p \\
\mathcal{M}, w \models \neg p &\iff w \notin \mathcal{M}p \\
\mathcal{M}, w \models s \vee t &\iff \mathcal{M}, w \models s \text{ or } \mathcal{M}, w \models t \\
\mathcal{M}, w \models s \wedge t &\iff \mathcal{M}, w \models s \text{ and } \mathcal{M}, w \models t \\
\mathcal{M}, w \models \Diamond s &\iff \exists v : w \rightarrow_{\mathcal{M}} v \text{ and } \mathcal{M}, v \models s \\
\mathcal{M}, w \models \Box s &\iff \forall v : w \rightarrow_{\mathcal{M}} v \text{ implies } \mathcal{M}, v \models s \\
\mathcal{M}, w \models \Diamond^* s &\iff \exists v : w \rightarrow_{\mathcal{M}}^* v \text{ and } \mathcal{M}, v \models s \\
\mathcal{M}, w \models \Box^* s &\iff \forall v : w \rightarrow_{\mathcal{M}}^* v \text{ implies } \mathcal{M}, v \models s
\end{aligned}$$

A model $\mathcal{M}$ **satisfies** (or is a **model of**) a formula $s$ if $\mathcal{M}, w \models s$ for some state $w \in |\mathcal{M}|$. A formula $s$ is **satisfiable** if $s$ has a model. The letter $A$ ranges over sets of formulas. We interpret sets of formulas conjunctively and write $\mathcal{M}, w \models A$ if $\mathcal{M}, w \models s$ for every formula $s \in A$. Moreover, $\mathcal{M}$ **satisfies** (or is a **model of**) $A$ if $\mathcal{M}, w \models A$ for some state $w \in |\mathcal{M}|$. Finally, a set $A$ is **satisfiable** if $\mathcal{M}, w \models A$ for some model $\mathcal{M}$ and some state $w \in |\mathcal{M}|$. We write $A; s$ for $A \cup \{s\}$.

Note that K* is not compact (consider $\Diamond^* \neg p, p, \Box p, \Box\Box p, \ldots$), and that the satisfiability problem for K* is EXPTIME-complete. The lower bound follows from Fischer and Ladner's proof for PDL (see [3], Theorem 6.52), and the upper bound is inherited from PDL [17], of which K* is a sublanguage.

A **formula universe** is a finite, nonempty set $\mathcal{F}$ of formulas satisfying the following properties:

- If $s \in \mathcal{F}$ and $t$ is a subformula of $s$, then $t \in \mathcal{F}$.
- If $\Diamond^* s \in \mathcal{F}$, then $\Diamond^+ s \in \mathcal{F}$.
- If $\Box^* s \in \mathcal{F}$, then $\Box^+ s \in \mathcal{F}$.

For every formula $s$ there exists a (least) formula universe $\mathcal{F}$ containing $s$ such that the cardinality of $\mathcal{F}$ is linear in the size of $s$. We will consider the following decision problem: Given a formula universe $\mathcal{F}$ and a formula $s \in \mathcal{F}$, decide whether $s$ is satisfiable. We assume $\mathcal{F}$ to be a global parameter and tacitly assume $s \in \mathcal{F}$ and $A \subseteq \mathcal{F}$ for every formula $s$ and every set of formulas $A$. Thus a set of formulas is always finite (since $\mathcal{F}$ is finite).

Finally, we fix some notations for binary relations. Let $\to \subseteq X \times X$.

$$\to^0 := \{ (x,x) \mid x \in X \} \qquad \to^* := \bigcup_{n \geq 0} \to^n$$

$$\to^{n+1} := \to \circ \to^n \qquad \to^+ := \to \circ \to^*$$

$$\mapsto := \{ (x,y) \in \to^* \mid \forall z: (y,z) \notin \to \}$$

For all relations shown above we use infix notation (i.e., $x \to^n y$ for $(x,y) \in \to^n$).

## 3 Demos

Decision procedures often construct syntactic descriptions of finite models. We introduce an abstract form of such descriptions and call them demos. The class of demos is adequate in that a formula is satisfiable if and only if it is supported by a demo. We use demos to establish the small model property and as the basis for the correctness proofs of our incremental tableau procedure. Technically, a demo is a set of clauses, where a clause is a set of formulas. Since we restrict formulas to a finite universe, there are only finitely many clauses and demos. This section builds on ideas in [13, 12] where the notion of support appears first.

We call formulas of the form $s \wedge t$ and $\Box^* s$ **conjunctive** and the remaining formulas **aconjunctive**. A **clause** is a nonempty set $C \subseteq \mathcal{F}$ of aconjunctive formulas. Our semantic definitions apply to clauses since they are sets of formulas. In particular, $\mathcal{M}, w \vDash C$ if and only if $\mathcal{M}, w \vDash s$ for every formula $s \in C$. A **literal** is a formula of the form $p$, $\neg p$, $\Diamond s$ or $\Box s$. A clause containing only literals is called an **$\alpha$-clause** provided it contains no complementary pair $p$ and $\neg p$. We define the **support relation** $C \rhd s$ between $\alpha$-clauses $C$ and formulas $s$ by induction on $s$:

$$\begin{aligned}
C \rhd s &\iff s \in C \quad \text{if } s \text{ is a literal} \\
C \rhd s \wedge t &\iff C \rhd s \text{ and } C \rhd t & C \rhd s \vee t &\iff C \rhd s \text{ or } C \rhd t \\
C \rhd \Diamond^* s &\iff C \rhd s \text{ or } C \rhd \Diamond^+ s & C \rhd \Box^* s &\iff C \rhd s \text{ and } C \rhd \Box^+ s
\end{aligned}$$

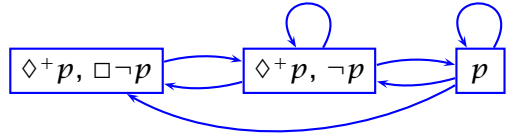

Figure 1: A demo and its transition relation

We say $C$ **supports** $s$ if $C \triangleright s$. Given an $\alpha$-clause $C$, the set $\{\, s \mid C \triangleright s \,\}$ satisfies the closure properties of a Hintikka set. Thus, we can see $\alpha$-clauses as a representation of certain Hintikka sets and the support relation as the concomitant membership relation. We write $C \triangleright A$ and say $C$ **supports** $A$ if $C \triangleright s$ for every $s \in A$.

**Lemma 3.1** Let $\mathcal{M}$ be a model such that $\mathcal{M}, w \vDash C$ and $C \triangleright s$. Then $\mathcal{M}, w \vDash s$.

**Proof** By induction on $s$. ∎

The **request** of a clause $C$ is $\mathcal{R}C := \{\, s \mid \Box s \in C \,\}$. Given a set $S$ of $\alpha$-clauses, we define the **transition relation** $\to_S \subseteq S \times S$ as follows: $C \to_S D \iff D \triangleright \mathcal{R}C$. A **demo** is a nonempty set $\mathcal{D}$ of $\alpha$-clauses such that:

1. If $\Diamond s \in C \in \mathcal{D}$, then $C \to_\mathcal{D} D \triangleright s$ for some $D \in \mathcal{D}$.
2. If $\Diamond^+ s \in C \in \mathcal{D}$, then $C \to_\mathcal{D}^+ D \triangleright s$ for some $D \in \mathcal{D}$.

Figure 1 shows a demo $\mathcal{D}$ consisting of three clauses $\{\Diamond^+ p, \Box\neg p\}$, $\{\Diamond^+ p, \neg p\}$, and $\{p\}$. The edges between the clauses depict the transition relation $\to_\mathcal{D}$.

A demo **supports** a formula $s$ or a set $A$ if it contains a clause that supports $s$ or $A$. For every nonempty set $S$ of $\alpha$-clauses we obtain a model $\mathcal{M}_S$ as follows: $|\mathcal{M}_S| = S$, $\to_{\mathcal{M}_S} = \to_S$, and $\mathcal{M}_S p = \{\, C \in S \mid p \in C \,\}$. We show that the model $\mathcal{M}_S$ obtained from a demo $S$ satisfies every formula supported by $S$.

**Lemma 3.2** If $C \triangleright \Box^* s$ and $C \to_S^n D$, then $D \triangleright \Box^* s$.

**Proof** By induction on $n$. ∎

**Lemma 3.3** Let $\mathcal{D}$ be a demo. Then $\forall s \, \forall C \in \mathcal{D}: C \triangleright s \implies \mathcal{M}_\mathcal{D}, C \vDash s$.

**Proof** By induction on $s$. Let $C \in \mathcal{D}$ and $C \triangleright s$. We show $\mathcal{M}_\mathcal{D}, C \vDash s$ by case analysis on $s$. All cases are straightforward except $s = \Box^* t$. Here the claim follows with Lemma 3.2. ∎

**Theorem 3.4** If $\mathcal{D}$ is a demo, then $\mathcal{M}_\mathcal{D}$ satisfies every formula that $\mathcal{D}$ supports.

**Proof** Immediate by Lemma 3.3.  ∎

Next we show that every satisfiable formula is supported by some demo. Let $\mathcal{M}$ be a model and $w \in |\mathcal{M}|$. Then $C_{\mathcal{M},w} := \{\, s \text{ literal} \mid \mathcal{M}, w \vDash s \,\}$ is an $\alpha$-clause. We show that $\mathcal{D}_{\mathcal{M}} := \{\, C_{\mathcal{M},w} \mid w \in |\mathcal{M}| \,\}$ is a demo.

**Lemma 3.5** Let $\mathcal{M}$ be a model and $\mathcal{M}, w \vDash s$. Then $C_{\mathcal{M},w} \rhd s$.

**Proof** By induction on $s$.  ∎

**Lemma 3.6** Let $\mathcal{M}$ be a model and $v \to_{\mathcal{M}} w$. Then $C_{\mathcal{M},v} \to_{\mathcal{D}_{\mathcal{M}}} C_{\mathcal{M},w}$.

**Proof** Follows with Lemma 3.5.  ∎

**Theorem 3.7** Let $\mathcal{M}$ be a model. Then $\mathcal{D}_{\mathcal{M}}$ is a demo. Moreover, $\mathcal{D}_{\mathcal{M}}$ supports a formula $s$ if and only if $\mathcal{M}$ satisfies $s$.

**Proof** First we show that $\mathcal{D}_{\mathcal{M}}$ is a demo. Let $\Diamond s \in C_{\mathcal{M},v} \in \mathcal{D}_{\mathcal{M}}$. We show that $C_{\mathcal{M},v} \to_{\mathcal{D}_{\mathcal{M}}} C_{\mathcal{M},w} \rhd s$ for some state $w$. Since $\Diamond s \in C_{\mathcal{M},v}$, there is a state $w$ such that $v \to_{\mathcal{M}} w$ and $\mathcal{M}, w \vDash s$. The claim follows with Lemmas 3.6 and 3.5. The second demo condition follows with a similar argument. Next we show that support by $\mathcal{D}_{\mathcal{M}}$ and satisfaction by $\mathcal{M}$ coincide.

Suppose $C_{\mathcal{M},w} \rhd s$. Since $\mathcal{M}, w \vDash C_{\mathcal{M},w}$, we have $\mathcal{M}, w \vDash s$ by Lemma 3.1.
Suppose $\mathcal{M}, w \vDash s$. Then $C_{\mathcal{M},w} \rhd s$ by Lemma 3.5. Thus $\mathcal{D}_{\mathcal{M}}$ supports $s$.  ∎

We formulate two immediate consequences of Theorems 3.4 and 3.7.

**Corollary 3.8** A formula is satisfiable if and only if it is supported by a demo.

**Corollary 3.9** Every satisfiable formula is satisfied by a finite model obtained from a demo.

Corollary 3.9 states a small model property for our logic. Corollary 3.8 gives us a purely syntactic decision method (support is decidable in polynomial time). It is easy to see that demos are closed under union. Hence there is a largest demo.

**Corollary 3.10** A formula is satisfiable if and only if it is supported by the largest demo.

## 4 Abstract Pruning

Given a set $\mathcal{U}$ of $\alpha$-clauses, it is straightforward to compute the largest demo contained in $\mathcal{U}$. One starts from $\mathcal{U}$ and stepwise deletes clauses in $\mathcal{U}$ that cannot occur in a demo contained in $\mathcal{U}$. We call this process abstract pruning. Abstract pruning terminates with the largest demo contained in $\mathcal{U}$. If we apply abstract pruning to the set of all $\alpha$-clauses, it computes the largest demo. This yields a decision procedure since a clause is satisfiable if and only if it is supported by the largest demo. Given that there are only exponentially many clauses, support can be decided in polynomial time, and a pruning step can be performed in exponential time, it is clear that the procedure can be realized so that it runs in deterministic exponential time (everything with respect to $|\mathcal{F}|$).

The sketched decision method is a reformulation of Pratt's [16] worst-case optimal decision method for PDL. Our $\alpha$-clauses replace the Hintikka sets used in Pratt's formulation. Proofs of the correctness of Pratt's method appear in [14, 10, 3]. These proofs argue semantically since they lack the notion of a demo. In contrast, our correctness argument delegates all semantic concerns to Corollary 3.10.

While abstract pruning will not be employed by our final decision procedure, it provides a helpful blueprint for the more elaborate pruning methods to be used.

Let $\mathcal{U}$ be a set of $\alpha$-clauses. We define the **abstract pruning relation** $\stackrel{AP}{\rightharpoonup}_\mathcal{U}$ as a binary relation on clause sets: $S \stackrel{AP}{\rightharpoonup}_\mathcal{U} S' \iff S \subsetneq S' \subseteq \mathcal{U}$ and there is a clause $C$ such that $S \cup \{C\} = S'$ and one of the following conditions holds:

1. There is some $\Diamond s \in C$ for which there is no $D$ such that $C \to_{\mathcal{U}\setminus S} D \triangleright s$.
2. There is some $\Diamond^+ s \in C$ for which there is no $D$ such that $C \to^+_{\mathcal{U}\setminus S} D \triangleright s$.

It is helpful to think of the clauses in $S$ and of $C$ as bad clauses that are pruned from $\mathcal{U}$. Note that $\stackrel{AP}{\rightharpoonup}_\mathcal{U}$ is a terminating relation (since $\mathcal{U} \subseteq 2^\mathcal{F}$ and $\mathcal{F}$ is finite).

**Lemma 4.1** Let $\mathcal{D} \subseteq \mathcal{U} \setminus S$ be a demo and $S \stackrel{AP}{\rightharpoonup}_\mathcal{U} S'$. Then $\mathcal{D} \subseteq \mathcal{U} \setminus S'$.

**Lemma 4.2** Let $S \stackrel{AP}{\rightharpoonup}_\mathcal{U} S'$. Then $\mathcal{U} \setminus S'$ is either empty or a demo.

**Theorem 4.3** Let $\emptyset \stackrel{AP}{\rightharpoonup}_\mathcal{U} S$. Then $\mathcal{U} \setminus S$ is either empty or the largest demo contained in $\mathcal{U}$. Moreover, $\mathcal{U}$ contains a demo if and only if $\mathcal{U} \setminus S \neq \emptyset$.

**Proof** Follows with Lemmas 4.1 and 4.2. ■

## 5 Tableaux

A decision procedure that starts from the set of all $\alpha$-clauses cannot be practical. Instead, we aim at a procedure that starts from a clause representing the input formula and that incrementally builds a graph tableau by adding clauses and links. A link is a triple $C\xi D$ where the annotation $\xi$ documents how the clause $D$ was obtained from the clause $C$. The procedure stops if the tableau determines the satisfiability or unsatisfiability of the input formula. Since the satisfiability or unsatisfiability of a formula can often be determined with a small tableau, the procedure provides for a practical prover. Figure 2 gives a first impression of how a graph tableau looks like. Graph tableaux seem to appear first in Pratt [17].

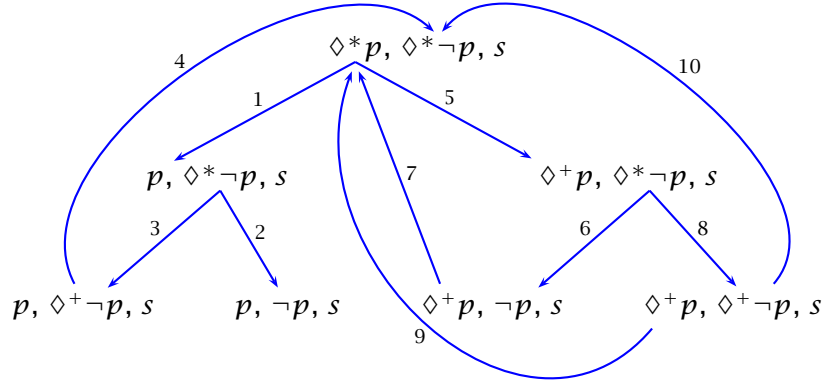

Figure 2: A complete tableau

Given a formula $s$, we define the **induced clause** $\lfloor s \rfloor$ by induction on $s$:

$$\begin{aligned}
\lfloor s \wedge t \rfloor &:= \lfloor s \rfloor \cup \lfloor t \rfloor \\
\lfloor \Box^* s \rfloor &:= \lfloor s \rfloor ; \Box^+ s \\
\lfloor s \rfloor &:= \{s\} \qquad \text{if } s \text{ aconjunctive}
\end{aligned}$$

Intuitively, the induced clause is obtained by decomposing all top-level conjunctions. The **induced clause** $\lfloor A \rfloor$ for a set $A$ of formulas is the union of the induced clauses for the formulas in $A$. We have $\mathcal{M}, w \vDash A$ if and only if $\mathcal{M}, w \vDash \lfloor A \rfloor$.

We partition the set of clauses into clashed clauses, $\alpha$-clauses, and $\beta$-clauses. A **clashed clause** is a clause that contains a pair $p$ and $\neg p$. A **$\beta$-clause** is a clause that is not clashed and that contains a formula of the form $s \vee t$ or $\Diamond^* s$.

We distinguish between $\alpha$-links and $\beta$-links. An **$\alpha$-link** is a triple $C\binom{\Diamond s}{s}D$ such that $C$ is an $\alpha$-clause, $\Diamond s \in C$, and $D = \lfloor \mathcal{R}C; s \rfloor$. A **$\beta$-link** is a triple $C\binom{u}{v}D$ such that $C$ is a $\beta$-clause, $u \in C$, $D = (C\setminus\{u\}) \cup \lfloor v \rfloor$, and one of the following conditions holds:

· $u = s \vee t$ and $v \in \{s, t\}$
· $u = \Diamond^* s$ and $v \in \{s, \Diamond^+ s\}$

A **tableau** $\mathcal{T}$ is a set of clauses and links such that the following conditions are satisfied:

1. If $C\binom{s}{t}D \in \mathcal{T}$, then $C$ and $D$ are in $\mathcal{T}$.
2. If $C\binom{s}{t}D$ and $C\binom{u}{v}E$ are $\beta$-links in $\mathcal{T}$, then $s = u$.

A graphical representation of a tableau appears in Figure 2. Links are represented as arrows. More concretely, a link $C\binom{s}{t}D$ is represented as an arrow that departs from the formula $s$ in $C$ and points to the formula $t$ in $D$. We see a tableau as a directed graph where the nodes are clauses and the edges are labeled with one or several binomial annotations. Given a tableau, we call a clause $D$ a **$\xi$-successor** of a clause $C$ if the tableau contains the link $C\xi D$. Note that a $\beta$-clause can have at most two successors in a tableau.

A **path** is a possibly empty sequence $(C_0\xi_0 C_1)(C_1\xi_1 C_2)\ldots(C_{n-1}\xi_{n-1}C_n)$ of links. We say that a **tableau contains a path** or that a **path is in a tableau** if the tableau contains every link of the path. Given a tableau $\mathcal{T}$, a clause $D$ is **reachable** from a clause $C$ if $\mathcal{T}$ contains a path whose first clause is $C$ and whose last clause is $D$. In the tableau in Figure 2, every clause is reachable from the topmost clause, which is a $\beta$-clause with two successors.

Let $\mathcal{T}$ and $\mathcal{T}'$ be two tableaux such that $\mathcal{T} \subseteq \mathcal{T}'$. Then we call $\mathcal{T}$ a **subtableau** of $\mathcal{T}'$ and $\mathcal{T}'$ a **supertableau** of $\mathcal{T}$. Moreover, we call a clause $C \in \mathcal{T}$ **expanded** if every link $C\xi D$ that appears in some supertableau of $\mathcal{T}$ already appears in $\mathcal{T}$. We call a tableau **complete** if each of its clauses is expanded. Note that the tableau in Figure 2 is complete. Given a tableau $\mathcal{T}$, one can compute a complete tableau that contains $\mathcal{T}$. Since only one disjunction in a $\beta$-clause can be expanded (condition (2) in the definition of tableaux), there may be different completions of a tableau.

Let $\mathcal{T}$ be a tableau and $S$ be a set of clauses. We write $\mathcal{T} \mid S$ for the largest subtableau of $\mathcal{T}$ containing only clauses from $S$, and $\mathcal{T}-S$ for the largest subtableau of $\mathcal{T}$ not containing a clause from $S$.

## 6 Evidence

The idea of a demo carries over to tableaux. We define a class of evident tableaux such that we obtain two properties:

1. The $\alpha$-clauses of an evident tableau comprise a demo supporting all clauses of the tableau.
2. Evidence of a tableau can be checked without checking support by just looking at the clauses and links of the tableau.

An evident tableau $\mathcal{T}$ must contain for every eventuality $\Diamond^* s \in C \in \mathcal{T}$ a fulfilling path that starts at $C$ and ends with a $\beta$-link annotated with $\binom{\Diamond^* s}{s}$. We call such paths runs. In the tableau in Figure 2, the eventuality $\Diamond^*\neg p$ in the topmost clause has a run consisting of the links 5 and 6. Furthermore, the eventuality $\Diamond^+ p$ in the rightmost clause has a run consisting of the links 9 and 1. Precise definitions follow.

We use the notation $\Diamond^\sigma s$ to denote eventualities of the form $\Diamond^* s$ or $\Diamond^+ s$. A **claim** $\Diamond^\sigma s \mid C$ is a pair of a clause $C$ and an eventuality $\Diamond^\sigma s \in C$. We define inductively what it means that a path $\pi$ is a **run for a claim** $\gamma$:

1. If $\pi = C\binom{\Diamond^* s}{s}D$, then $\pi$ is a run for $\Diamond^* s \mid C$.
2. If $\pi = (C\binom{\Diamond^+ s}{\Diamond^* s}D)\pi'$ and $\pi'$ is a run for $\Diamond^* s \mid D$, then $\pi$ is a run for $\Diamond^+ s \mid C$.
3. If $\pi = (C\binom{s}{t}D)\pi'$, $C$ is a $\beta$-clause, and $\pi'$ is a run for $\Diamond^\sigma s \mid D$, then $\pi$ is a run for $\Diamond^\sigma s \mid C$.

A tableau $\mathcal{T}$ is **evident** if it satisfies the following conditions:

1. $\mathcal{T}$ contains no clashed clause.
2. If $\Diamond s \in C \in \mathcal{T}$ and $C$ is an $\alpha$-clause, then $\mathcal{T}$ contains a link $C\binom{\Diamond s}{s}D$.
3. If $C \in \mathcal{T}$ is a $\beta$-clause, then $C$ has at least one successor in $\mathcal{T}$.
4. If $\Diamond^\sigma s \in C \in \mathcal{T}$, then $\mathcal{T}$ contains a run for $\Diamond^\sigma s \mid C$.

A clause $C$ is **evident in a tableau** $\mathcal{T}$ if $C$ is contained in an evident subtableau of $\mathcal{T}$. We use $\mathcal{E}^\mathcal{T}$ to denote the set of all clauses that are evident in $\mathcal{T}$. Note that the union of evident tableaux is an evident tableau. Hence $\mathcal{T} \mid \mathcal{E}^\mathcal{T}$ is the largest evident subtableau of $\mathcal{T}$.

Consider the tableau in Figure 2. The largest evident evident subtableau of this tableau is obtained by deleting link 2 and the clashed clause it points to. The situation changes if we delete link 5. Then the largest evident subtableau is empty. The example tells us that non-evident clauses may be satisfiable.

**Proposition 6.1** Let $\mathcal{U}$ and $\mathcal{T}$ be tableaux such that $\mathcal{U} \subseteq \mathcal{T}$. Then $\mathcal{E}^\mathcal{U} \subseteq \mathcal{E}^\mathcal{T}$.

**Proposition 6.2** If $C\xi D$ is a $\beta$-link and $E \rhd D$, then $E \rhd C$.

**Lemma 6.3** Let $\mathcal{T}$ be an evident tableau and $C \in \mathcal{T}$ be a clause. Then there exists an $\alpha$-clause $D \in \mathcal{T}$ that supports $C$.

**Proof** By induction on the sum of the sizes of the non-literal formulas in $C$ using Proposition 6.2. ∎

**Theorem 6.4** The $\alpha$-clauses of a nonempty evident tableau comprise a demo that supports every clause of the tableau.

**Corollary 6.5** If a clause is evident in a tableau, then it is satisfiable.

**Proof**  Follows with Theorems 6.4 and 3.4.  ∎

Next, we lift Theorem 3.7 from demos to evident tableaux. Let $\mathcal{T}$ be a tableau and $\mathcal{M}$ be a model. We define $\mathcal{T} \mid \mathcal{M} := \mathcal{T} \mid \{\, C \in \mathcal{T} \mid \mathcal{M} \text{ satisfies } C \,\}$.

**Lemma 6.6**  Let $\mathcal{T}$ be a complete tableau, $\mathcal{M}$ be a model, $C \in \mathcal{T} \mid \mathcal{M}$ be a clause, $\diamond^\sigma s \in C$, $\mathcal{M}, w \vDash C$, $w \to^n_\mathcal{M} v$, $\mathcal{M}, v \vDash s$, and $n > 0$ if $\sigma = +$. Then $\diamond^\sigma s \mid C$ has a run in $\mathcal{T} \mid \mathcal{M}$.

**Proof**  By lexicographic induction on $n$ and the sum of the sizes of the non-literal formulas in $C$.  ∎

**Lemma 6.7**  Let $\mathcal{T}$ be a complete tableau and $\mathcal{M}$ be a model. Then $\mathcal{T} \mid \mathcal{M}$ is an evident subtableau of $\mathcal{T}$.

**Proof**  Follows with Lemma 6.6.  ∎

**Theorem 6.8**  Let $\mathcal{T}$ be a complete tableau. Then a clause $C \in \mathcal{T}$ is satisfiable if and only if $C$ is evident in $\mathcal{T}$.

**Proof**  Follows with Lemma 6.7 and Corollary 6.5.  ∎

**Corollary 6.9**  Let $\mathcal{T}$ be a complete tableau and $\lfloor s \rfloor \in \mathcal{T}$. Then $s$ is satisfiable if and only if $\lfloor s \rfloor$ is evident in $\mathcal{T}$.

## 7 Eager Pruning

The idea of pruning carries over from demos to tableaux. Given a tableau, one stepwise marks clauses that cannot appear in an evident subtableau. We call this process eager pruning. When eager pruning terminates, exactly the non-evident clauses of the tableau are marked. Eager pruning gives us a worst-case optimal decision method: Given a formula $s$, construct a complete tableau containing $\lfloor s \rfloor$ and run eager pruning. Then $s$ is satisfiable if and only if $\lfloor s \rfloor$ has not been marked. A similar method appears first in Pratt [17].

Let $\mathcal{T}$ be a tableau. We define the **eager pruning relation** $\xrightarrow{\text{EP}}_\mathcal{T}$ as a binary relation on clause sets: $S \xrightarrow{\text{EP}}_\mathcal{T} S' \iff S \subsetneq S' \subseteq \mathcal{T}$ and there is a clause $C$ such that $S \cup \{C\} = S'$ and one of the following conditions holds:

1. $C$ is clashed.
2. $C$ is an $\alpha$-clause that is not expanded in $\mathcal{T}$.
3. $C$ is an $\alpha$-clause that has a successor in $\mathcal{T}$ that is in $S$.
4. $C$ is a $\beta$-clause all of whose successors in $\mathcal{T}$ are in $S$.

5. $C$ contains an eventuality $s$ such that every run for $s \mid C$ in $\mathcal{T}$ contains a clause in $S$.

Note that $\overset{\text{EP}}{\mapsto}_{\mathcal{T}}$ is a terminating relation.

**Lemma 7.1 (Soundness)** Let $S \subseteq \mathcal{T} \setminus \mathcal{E}^{\mathcal{T}}$ and $S \overset{\text{EP}}{\mapsto}_{\mathcal{T}} S'$. Then $S' \subseteq \mathcal{T} \setminus \mathcal{E}^{\mathcal{T}}$.

**Lemma 7.2** Let $S \overset{\text{EP}}{\mapsto}_{\mathcal{T}} S'$. Then $\mathcal{T} - S'$ is an evident tableau.

**Theorem 7.3** Let $\mathcal{T}$ be a tableau, $\mathcal{R} \overset{\text{EP}}{\mapsto}_{\mathcal{T}} S$, and $\mathcal{R}$ be a set of unsatisfiable clauses. Then $\mathcal{E}^{\mathcal{T}} = \{ C \in \mathcal{T} \mid C \notin S \}$.

**Proof** Follows with Lemmas 7.1 and 7.2. ∎

Consider the tableau in Figure 2. If we apply eager pruning to the tableau, it marks the clashed clause and no other clause. The situation changes if we apply eager pruning to the tableau with link 5 removed. Then all clauses of the tableau are marked as non-evident. This is the case since all but one eventuality for $\neg p$ lose their runs. The remaining two clauses then lack runs for the eventuality $\diamond^+ p$.

## 8 Cautious Pruning

We define cautious pruning as a constrained form of eager pruning that marks unsatisfiable clauses. We base cautious pruning on the following consequence of Theorem 6.8.

**Corollary 8.1** Let $\mathcal{T}$ be a tableau. Then a clause $C \in \mathcal{T}$ is unsatisfiable if and only if there is no complete supertableau of $\mathcal{T}$ in which $C$ is evident.

Cautious pruning will mark a clause in a tableau if it is clear that the clause cannot become evident by adding further links and clauses. To account for eventualities, we need the notion of a plan. A plan is a path in a tableau that is either a run or a partial run that ends with a non-expanded clause. If a claim has a plan, it may have a run in a completion of the tableau. However, if a claim has no plan, it cannot have a run in any completion of the tableau.

We define inductively what it means that a path $\pi$ in a tableau $\mathcal{T}$ is a **plan for a claim $\gamma$ in $\mathcal{T}$**:

1. If $\diamond^+ s \in C \in \mathcal{T}$, $C$ is an $\alpha$-clause, and $C$ has no $\binom{\diamond^+ s}{\diamond^* s}$-successor in $\mathcal{T}$, then the empty path is a plan for $\diamond^+ s \mid C$ in $\mathcal{T}$.
2. If $\diamond^\sigma s \in C \in \mathcal{T}$ and $C$ is a $\beta$-clause that is not expanded in $\mathcal{T}$, then the empty path is a plan for $\diamond^\sigma s \mid C$ in $\mathcal{T}$.

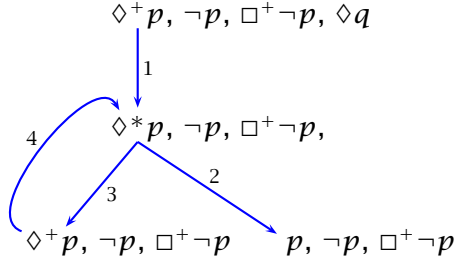

Figure 3: An incomplete tableau

3. If $\pi = C\binom{\diamond^* s}{s}D$, then $\pi$ is a plan for $\diamond^* s \mid C$ in $\mathcal{T}$.
4. If $\pi = (C\binom{\diamond^+ s}{\diamond^* s}D)\pi'$ and $\pi'$ is a plan for $\diamond^* s \mid D$ in $\mathcal{T}$, then $\pi$ is a plan for $\diamond^+ s \mid C$ in $\mathcal{T}$.
5. If $\pi = (C\binom{s}{t}D)\pi'$, $C$ is a $\beta$-clause, and $\pi'$ is a plan for $\diamond^\sigma s \mid D$ in $\mathcal{T}$, then $\pi$ is a plan for $\diamond^\sigma s \mid C$ in $\mathcal{T}$.

**Lemma 8.2** Let $\mathcal{T}$ be a tableau and $\gamma$ be a claim whose clause is in $\mathcal{T}$. Then:

1. Every run for $\gamma$ in $\mathcal{T}$ is a plan for $\gamma$ in $\mathcal{T}$.
2. Every plan for $\gamma$ in $\mathcal{T}$ containing only clauses expanded in $\mathcal{T}$ is a run for $\gamma$.
3. If $\gamma$ has a plan $\pi$ in a supertableau of $\mathcal{T}$, then it has a plan in $\mathcal{T}$ that is a prefix of $\pi$.

Let $\mathcal{T}$ be a tableau. We define the **cautious pruning relation** $\xrightarrow{\text{CP}}_{\mathcal{T}}$ as a binary relation on clause sets: $S \xrightarrow{\text{CP}}_{\mathcal{T}} S' \iff S \subsetneq S' \subseteq \mathcal{T}$ and there is a clause $C$ such that $S \cup \{C\} = S'$ and one of the following conditions holds:

1. $C$ is clashed.
2. $C$ is an $\alpha$-clause that has a successor in $\mathcal{T}$ that is in $S$.
3. $C$ is a $\beta$-clause that is expanded in $\mathcal{T}$ and all successors of $C$ in $\mathcal{T}$ are in $S$.
4. $C$ contains an eventuality $s$ such that every plan for $s \mid C$ in $\mathcal{T}$ contains a clause in $S$.

Consider the tableau in Figure 3. Cautious pruning will mark all clauses of the tableau since they are either clashed or contain an eventuality that has no plan. Note that the tableau is incomplete since the topmost clause has no successor for the formula $\diamond q$. If we replace $q$ with a complex formula, completing the tableau may be expensive.

**Proposition 8.3** Let $\mathcal{T}$ be a tableau. Then $\xrightarrow{\text{CP}}_{\mathcal{T}} \subseteq \xrightarrow{\text{EP}}_{\mathcal{T}}$. Moreover, $\xrightarrow{\text{CP}}_{\mathcal{T}}$ is terminating and confluent. Hence there exists a unique set $\mathcal{R}$ such that $\emptyset \xrightarrow{\text{CP}}_{\mathcal{T}} \mathcal{R}$.

**Proof** We have $\xrightarrow{\text{CP}}_{\mathcal{T}} \subseteq \xrightarrow{\text{EP}}_{\mathcal{T}}$ since every run is a plan. Cautious pruning is terminating since there are only finitely many clauses, and confluent since it satisfies the diamond property [2]. ∎

Given a tableau $\mathcal{T}$, we use $\mathcal{R}^{\mathcal{T}}$ to denote the unique set $\mathcal{R}$ of clauses such that $\emptyset \xrightarrow{\text{CP}}_{\mathcal{T}} \mathcal{R}$. We say that that a clause $C$ is **refuted in** $\mathcal{T}$ if $C \in \mathcal{R}^{\mathcal{T}}$.

**Proposition 8.4** Let $S \xrightarrow{\text{CP}}_{\mathcal{T}} S'$ and $S \subseteq \mathcal{R}^{\mathcal{T}}$. Then $S' \subseteq \mathcal{R}^{\mathcal{T}}$.

**Proposition 8.5** If $\mathcal{T}$ is a complete tableau, then $\xrightarrow{\text{CP}}_{\mathcal{T}} = \xrightarrow{\text{EP}}_{\mathcal{T}}$.

**Lemma 8.6** Let $\mathcal{U}$ and $\mathcal{T}$ be tableaux such that $\mathcal{U} \subseteq \mathcal{T}$. Then $\xrightarrow{\text{CP}}_{\mathcal{U}} \subseteq \xrightarrow{\text{CP}}_{\mathcal{T}}$ and $\mathcal{R}^{\mathcal{U}} \subseteq \mathcal{R}^{\mathcal{T}}$.

**Proof** Follows with Lemma 8.2 (3). ∎

Lemma 8.6 states that cautious pruning is monotone with respect to the underlying tableau. This is not the case for eager pruning (consider the tableau $\{\{\Diamond p\}\}$).

**Lemma 8.7** Let every clause in $S$ be unsatisfiable and $S \xrightarrow{\text{CP}}_{\mathcal{T}} S'$. Then every clause in $S'$ is unsatisfiable.

**Proof** Consider the 4 cases in the definition of $\xrightarrow{\text{CP}}_{\mathcal{T}}$. Except for the 4th case the claim is obvious. Now let $C \in \mathcal{T}$ be a clause that contains an eventuality $s$ such that every plan for $s \mid C$ in $\mathcal{T}$ contains a clause in $S$. Furthermore, let $\mathcal{T}'$ be a complete supertableau of $\mathcal{T}$. Suppose $C$ is satisfiable. By Theorem 6.8, $C$ is evident in $\mathcal{T}'$. Hence there is an evident subtableau of $\mathcal{T}'$ that contains a run $\pi$ for $s \mid C$. Since the clauses in $S$ are unsatisfiable, it follows with Corollary 6.5 that $\pi$ contains no clause in $S$. By Lemma 8.2 (3), some prefix of $\pi$ is a plan for $s \mid C$ in $\mathcal{T}$. Contradiction. ∎

**Theorem 8.8 (Soundness)** Every clause in $\mathcal{R}^{\mathcal{T}}$ is unsatisfiable.

**Proof** Follows from Lemma 8.7 ∎

**Lemma 8.9 (Progress)** Let $\mathcal{R}^{\mathcal{T}} \xrightarrow{\text{EP}}_{\mathcal{T}} S$ and $\mathcal{R}^{\mathcal{T}} \subsetneq S$. Then $S \setminus \mathcal{R}^{\mathcal{T}}$ contains a clause that is not expanded in $\mathcal{T}$.

Input:   a formula $s \in \mathcal{F}$
Variables:   $\mathcal{T} := \{\lfloor s \rfloor\}$,   $\mathcal{R} := \emptyset$,   $\mathcal{E} := \emptyset$
Invariants:

- $\mathcal{T}$ is a tableau such that every clause is reachable from $\lfloor s \rfloor$
- $\mathcal{R} \subseteq \mathcal{R}^\mathcal{T}$ and $\mathcal{E} \subseteq \mathcal{E}^\mathcal{T}$

while $\lfloor s \rfloor \notin \mathcal{R} \cup \mathcal{E}$ do one of the following:

- Grow $\mathcal{T}$ by adding a link to a clause in $\mathcal{T} \setminus (\mathcal{R} \cup \mathcal{E})$
- Grow $\mathcal{R}$ with cautious pruning, that is, $\mathcal{R} := \mathcal{R}'$ where $\mathcal{R} \overset{\text{CP}}{\mapsto}_\mathcal{T} \mathcal{R}'$
- Grow $\mathcal{E}$ with eager pruning, that is, $\mathcal{E} := \{\, C \in \mathcal{T} \mid C \notin S \,\}$ where $\mathcal{R} \overset{\text{EP}}{\mapsto}_\mathcal{T} S$

Output:   SAT if $\lfloor s \rfloor \in \mathcal{E}$ and UNSAT otherwise

Figure 4: Incremental decision procedure IDP

**Proof**  By Theorem 7.3 and Theorem 8.8 we know that the tableau $\mathcal{T}-S$ is evident and that the tableau $\mathcal{T}-\mathcal{R}^\mathcal{T}$ is not evident. Hence there is a clause $C \in S \setminus \mathcal{R}^\mathcal{T}$ that violates an evidence condition for $\mathcal{T}-\mathcal{R}^\mathcal{T}$.

We obtain the claim by contradiction. Suppose every clause in $S \setminus \mathcal{R}^\mathcal{T}$ is expanded in $\mathcal{T}$. Then $C$ is expanded in $\mathcal{T}$. Since $C$ violates an evidence condition for $\mathcal{T}-\mathcal{R}^\mathcal{T}$, there must be an eventuality $s \in C$ such that $s \mid C$ has no run in $\mathcal{T}-\mathcal{R}^\mathcal{T}$. Since $C \notin \mathcal{R}^\mathcal{T}$, the claim $s \mid C$ has a plan in $\mathcal{T}$ that does not contain a clause in $\mathcal{R}^\mathcal{T}$. By Lemma 8.2 (3), $s \mid C$ then has a plan in $\mathcal{T}-\mathcal{R}^\mathcal{T}$. Since every clause in $S \setminus \mathcal{R}^\mathcal{T}$ is expanded and every eventuality in $\mathcal{T}-S$ has a run in $\mathcal{T}-S$ (since $\mathcal{T}-S$ is evident), $s \mid C$ has a run in $\mathcal{T}-\mathcal{R}^\mathcal{T}$ (follows with Lemma 8.2 (2)). Contradiction.  ∎

## 9  An Incremental Decision Procedure

We now have everything in place to formulate an incremental decision procedure. Starting from an input formula $s$, the procedure grows a tableau $\mathcal{T}$ and two clause sets $\mathcal{R} \subseteq \mathcal{T}$ and $\mathcal{E} \subseteq \mathcal{T}$ such that all clauses in $\mathcal{R}$ are refuted and all clauses in $\mathcal{E}$ are evident in $\mathcal{T}$. The set $\mathcal{R}$ is grown by cautious pruning, and the set $\mathcal{E}$ is grown by eager pruning (taking complements). The procedure stops once the initial clause $\lfloor s \rfloor$ appears in $\mathcal{R}$ or $\mathcal{E}$. Figure 4 gives a precise formulation of the procedure, which we name IDP.

We argue the correctness of IDP as follows:

1. Preservation of the invariant for $\mathcal{R}$ follows from Lemma 8.6 and Proposition 8.4. Preservation of the invariant for $\mathcal{E}$ follows from Theorems 8.8 and 7.3, and Proposition 6.1.

2. That the loop can perform at least one of the three possible steps follows from the loop condition, the invariants, and Lemma 8.9.

3. The loop terminates since each iteration grows one of the sets $\mathcal{T}$, $\mathcal{R}$ and $\mathcal{E}$, and each of these sets is bounded by the finite formula universe $\mathcal{F}$.

4. Correctness of the output follows from the negated loop condition, the invariants, Corollary 6.5, and Theorem 8.8.

IDP runs in deterministic exponential time with respect to the size of $s$. To see this, note that every traversal of the loop adds at least one element to one of the sets $\mathcal{T}$, $\mathcal{R}$, and $\mathcal{E}$, and that the cardinality of each of these sets is exponentially bounded in $|\mathcal{F}|$. Moreover, each of the three possible actions of the loop can be realized in polynomial time with respect to the size of $\mathcal{T}$, $\mathcal{R}$, and $\mathcal{E}$.

We call a tableau $\mathcal{T}$ a **certificate** for a formula $s$ and say that $\mathcal{T}$ **determines** $s$ if $\lfloor s \rfloor \in \mathcal{R}^{\mathcal{T}} \cup \mathcal{E}^{\mathcal{T}}$ and every clause in $\mathcal{T}$ is reachable from $\lfloor s \rfloor$. We speak of a **negative certificate** if $\lfloor s \rfloor \in \mathcal{R}^{\mathcal{T}}$, and of a **positive certificate** if $\lfloor s \rfloor \in \mathcal{E}^{\mathcal{T}}$. For example, the tableau in Figure 3 is a negative certificate for $\Diamond^+ p \wedge \Box^* \neg p \wedge \Diamond q$, and the tableau in Figure 2 is a positive certificate for $\Box^*(\Diamond^* p \wedge \Diamond^* \neg p)$. Note that it can be decided in polynomial time whether a tableau is a positive or negative certificate for a formula (with respect to the size of the tableau).

**Proposition 9.1** IDP always terminates with a tableau that determines the input formula. Thus every formula has a certificate.

**Proof** Follows from the invariants and the negated loop condition. ∎

Vice versa, let $\mathcal{T}$ determine $s$. Then IDP can first construct the tableau $\mathcal{T}$ and then do the necessary pruning to establish $\mathcal{T}$ as a certificate of $s$. Thus IDP can be efficient for formulas with small certificates. Of course, the question remains how IDP selects the right tableau expansion steps. Given the complexity of the problem, there will be no perfect answer. Nevertheless, it is a good idea to grow $\mathcal{R}$ and $\mathcal{E}$ eagerly so that unnecessary expansions are avoided. Moreover, one can show that it is unnecessary to expand clauses that can only be reached from the input formula by a path that passes through a clause in $\mathcal{R}$ or $\mathcal{E}$.

## 10 Branching Tableau Search

Provers for modal logic often employ branching and backtracking to search for an evident tableau. Transferred to our setting, branching search branches on $\beta$-clauses (admitting only one departing link per $\beta$-clause and branch) and stops as soon as a branch contains a clashed clause or a bad loop. Consider the complete tableau in Figure 2 for illustration. Branching search unrolls this tableau into four branches, containing the links $\{1, 2\}$, $\{1, 3, 4\}$, $\{5, 6, 7\}$, and $\{5, 8, 9, 10\}$, respectively. The branch $\{1, 2\}$ is failed (i.e., closed) since it contains a clashed clause, and the other branches are failed since they contain bad loops leaving eventualities unfulfilled. However, the tableau in Figure 2 becomes evident if we delete the clashed clause. Thus the example tells us that branching search is incomplete for our tableau system.

There are branching tableau systems that are complete for modal logic with eventualities. The systems of Baader [1] and De Giacomo and Massacci [4] employ prefixes and blocking, and a recent system of the authors [13] employs $\alpha$-clauses and a DNF operator replacing $\beta$-clauses. Although branching tableau search is not worst-case optimal for modal logic with eventualities, it may still perform well in practice due to its incrementality. Moreover, the branching system in [13] accommodates nominals without sacrificing incrementality. So far it is open whether one can have an incremental and worst-case optimal decision procedure for modal logic with nominals and eventualities.